\documentclass[12pt]{iopart}
\usepackage{epsfig,graphicx}

\newcommand{\piz} {\ensuremath{\pi^0\,}}
 
\def\pt#1{\ensuremath{p_{\rm T#1}}}

\newcommand{\xe} {\ensuremath{x_{\rm E}\,}}

\newcommand{\vs}{$\sqrt{s}$}

\newcommand{\dphi}{\mbox{$\Delta \phi\,$}}

\newcommand{\zt}{\mbox{$z_{\mathrm{t}}\,$}}
\RequirePackage{lineno}
\begin{document}
\title[]{High \pt{} particle correlations in pp collisions at LHC/ALICE}
 
\author{Yaxian Mao$^{1,2}$, for the ALICE collaboration}

\address{$^{1}$ Key Laboratory of Quark \& Lepton Physics~(Huazhong Normal University),  Ministry of Education, Wuhan 430079, China \\
$^{2}$ Laboratoire de Physique Subatomique et de Cosmologie, Grenoble 38026, France}
\ead{Yaxian.Mao@cern.ch}
\begin{abstract}
Two-particle correlation triggered by high-\pt{} particles 
allows us to study hard scattering phenomena when full jet reconstruction is challenging. 
An analysis of the first ALICE pp data where charged and neutral particles isolated or not are used as trigger particles is presented.
The two-particle correlation between the trigger ($t$) and the associate ($a$) particles is studied as a function of the imbalance parameter \xe=-$\vec{p}_{T_{a}} \cdot \vec{p}_{T_{t}}/\mid \vec{p}_{T_{t}}\mid ^{2}$ and interpreted in terms of jet fragmentation function.
\end{abstract}

\linenumbers
Two-particle correlation measurements provide a suitable tool to study the jet properties in proton-proton and heavy-ion collisions~\cite{PHENIX1, PHENIX2} where the full jet reconstruction is challenging and even impossible for jet-\pt{} below about 50~GeV/c. 
Such a measurement can be further improved if the trigger particle is isolated. Ultimately, selecting a direct photon as the trigger particle would provide the optimum choice since the direct photon sets the reference of the  $2\rightarrow 2$ type of hard scattering kinematics and balances the jet emitted in the opposite azimuthal direction. In addition, while purely hadronic observables are strongly biased with respect to surface emission, photons can sample the entire collision volume. However, identifying direct photons is quite challenging because of their scarcity and the overwhelming contribution of hadron decay-photons, mainly from \piz. Applying isolation criteria (select trigger particles spatially isolated) is the traditional approach to enrich the data sample with hadrons sampling a large fraction of the parton momentum from which they fragment.

We have studied the two-particle correlation with the ALICE experiment~\cite{ALICE} at LHC in proton-proton collisions at \vs$=7$~TeV. 
The experimental technique consists in tagging events with a leading trigger and measuring the distribution of charged hadrons associated to this leading trigger from the same event. 
Such a measurement requires 
to reconstruct charged tracks and
identify neutral particles with good momentum/energy resolution. In ALICE, the
electromagnetic calorimeters, PHOS ($|\Delta\eta|<0.12$ and $\Delta
\phi$ =100$^o$) and EMCAL ($|\Delta\eta|<0.7$ and $\Delta \phi$
=100$^o$), allow us to measure and identify photons and neutral mesons with high
efficiency and resolution. Particles in the calorimeters are detected as clusters of adjacent hit calorimeter cells.
The Central Tracking System (ITS and TPC), covering the pseudo-rapidity
$-0.9 \leq \eta \leq +0.9$ and the full azimuth~\cite{ALICE}, is used for charged track measurements, and contributes to the single particle jet events identification by applying the isolation technique. 
Two different types of trigger particle have been selected for the correlation study: (i)  charged triggers are selected as the charged track with the highest transverse momentum among all the tracks in the event, (ii)   neutral cluster triggers are defined as the calorimeter cluster with the highest energy among all detected clusters and among all the charged tracks detected in the same trigger hemisphere. 
At this early stage of the analysis, no particle identification has been applied yet on the calorimeters data, thus the cluster sample may contain a sizable fraction of charged particles which develop a shower in the calorimeters or high-$p_{T}$ $\pi^{0}$ from which two decay photons merge into a single cluster. 
A Monte-Carlo study based on the PYTHIA~\cite{PYTHIA} event generator indicates that the dominant contribution to the selected neutral cluster triggers is due to merged~\piz in the \pt{} region which could be statistically reached with the data available for our study. 
The results which will be discussed are the azimuthal correlation between the two particles, and the per-trigger conditional yield of charged hadrons as a function of the imbalance variable \xe= -$\frac{\vec{p}_{T_{t}} \cdot \vec{p}_{\mathrm T_{a}}} {\mid \vec{p}_{\mathrm T_{t}} \mid ^{2}}$, where $\vec{p}_{\mathrm T_{t, a}}$ denote the amplitude of the transverse momenta of the trigger and associate particle, respectively. 

\begin{figure}[h]
\begin{minipage}{19pc}
\includegraphics[width=0.9\textwidth,clip]{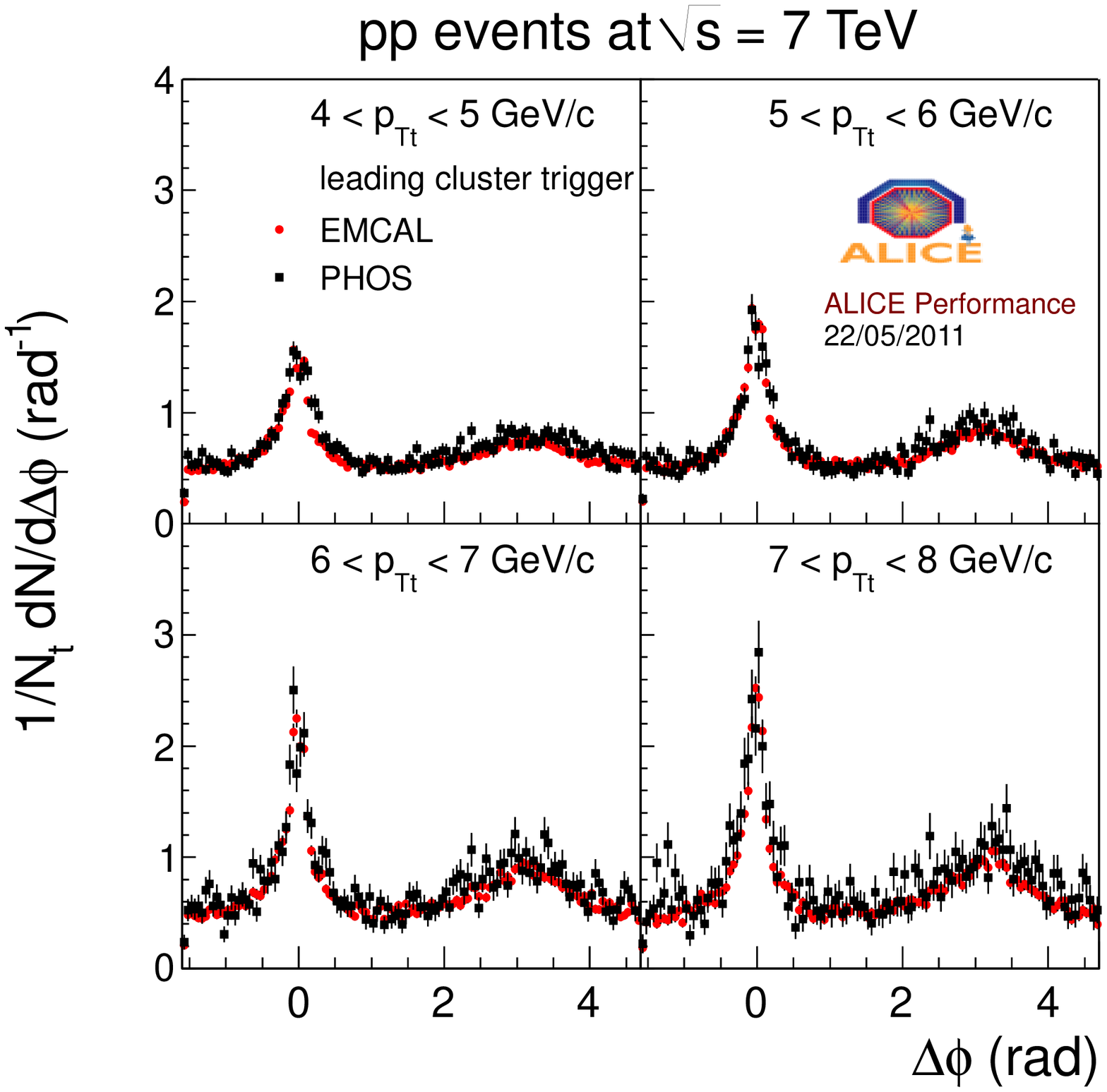}
\caption{\em Two-particle correlation distribution as a function of the relative azimuthal angle between the cluster trigger and  associate charged particles  \dphi$=\phi_{\mathrm{t}}-\phi_{\mathrm{a}}$ in pp collisions at \vs~= 7 TeV.}
\label{fig:ClusterDphi}
\end{minipage}
\begin{minipage}{19pc}
\includegraphics[width=0.9\textwidth,clip]{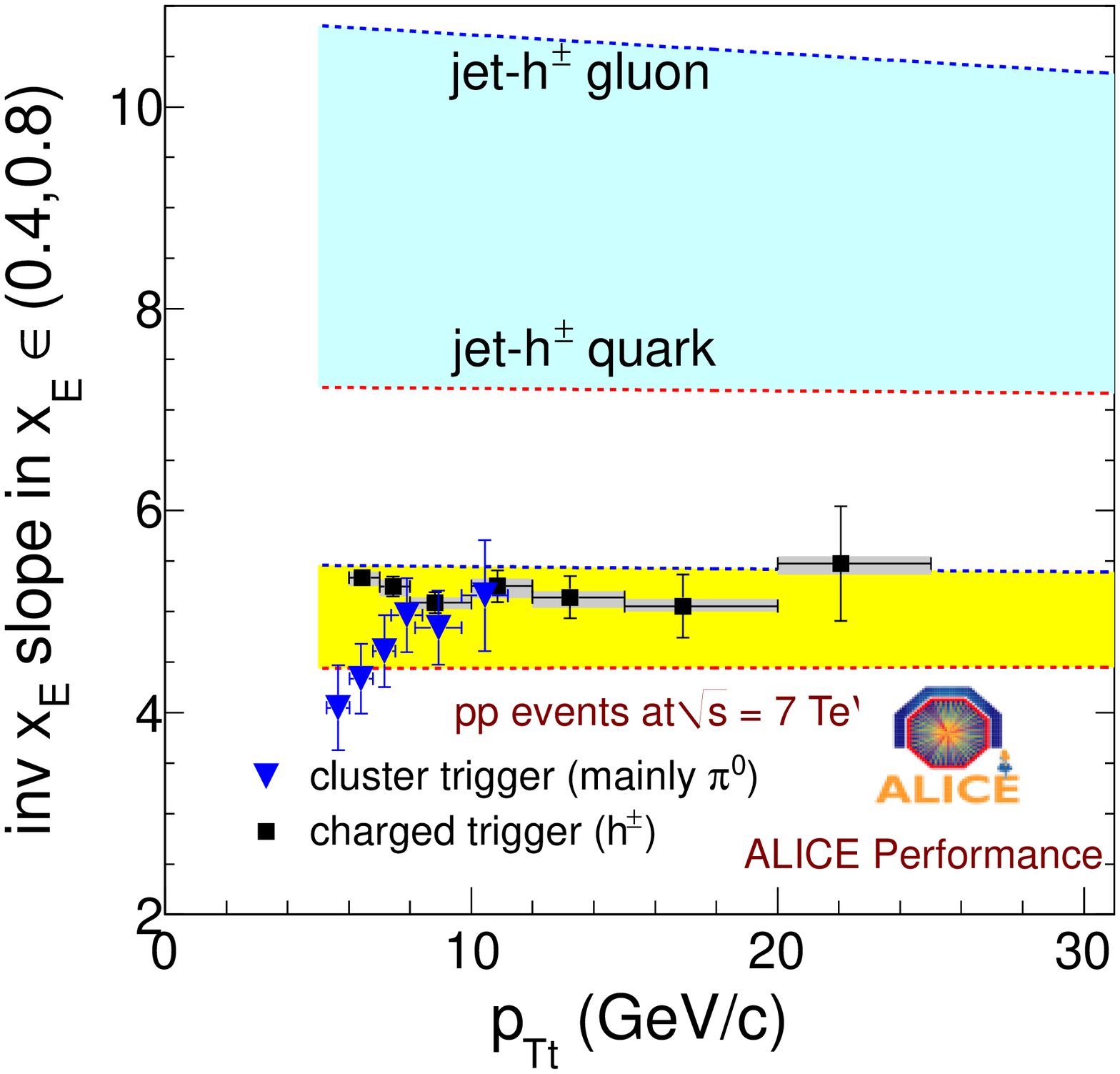}
\caption{\em Extracted inverse \xe slope as a function of the mean trigger \pt{t}~ for charged and neutral triggers in pp collisions at \vs~= 7 TeV.}
\label{fig:xeSlope}
\end{minipage}
\end{figure}

The azimuthal correlation  ($\Delta \phi = \phi_{\mathrm{t}} - \phi_{\mathrm{a}}$) between the trigger particle and the associate particles with transverse momentum threshold of \pt{}~$>1~$GeV/$c$ and $|\eta| < 0.8$ for different \pt{} trigger bins is shown in Fig.~\ref{fig:ClusterDphi}. 
The resulting \dphi distribution is finally corrected for the two-particle pair efficiency. The structure of the azimuthal correlations show consistent results independent on the trigger type and of the calorimeter for the cluster triggers (Fig.~\ref{fig:ClusterDphi}). 
The main feature of these distributions is the typical  2-jets structure with a near side (\dphi$=0$) and away side (\dphi$=\pi$) peak. 
The peaks become stronger when the trigger-\pt{} increases reflecting the increasing multiplicity of fragmented hadrons for increasing jet energies (on average the leading particle of a jet carries about 50\% of its energy at LHC energy regime).
The 2-jet structure sits on top of a flat background which originates from the correlation of the trigger particle with particles from the underlying event (all particles which do not originate from a hard scattering process).  

The per-trigger conditional yield as a function of the variable \xe
is used to study the away side jet fragmentation function. The accuracy of this description depends on how well the trigger particle momentum approximates the jet momentum. 
The measured \xe distribution contains contributions from away side jet fragmentation signal as well as background. 
To estimate this background, one assumes that the associate particles distribution from underlying events is isotropic in the full azimuth angle. Therefore, the background level can be calculated from the transverse azimuthal region where the jet signal is minimum. 
The resulting \xe signal distributions after underlying event background subtraction for different \pt{t} bins exhibits remarkable uniformity of the slope. 
In order to quantity this feature, an exponential function (dN/d\xe $\propto Ce^{-n\cdot x_\mathrm{E}}$) is fitted to the \xe distribution in the range $0.4~<$ \xe~$<~0.8$, the fitting parameter $n$ stands for the inverse \xe~slope. 
Figure~\ref{fig:xeSlope} shows the evolution of the inverse \xe~slope as a function of the mean trigger \pt{} obtained from ALICE data. 
The slopes extracted using charged and neutral triggers approach to each other with rising \pt{}. 
This behavior is expected as high \pt{} cluster triggers are mainly merged \piz while low \pt{} clusters are 
dominated by single decay photon clusters from \piz. 

\begin{figure}[h]
\centering
\begin{minipage}{19pc}
\includegraphics[width=0.85\textwidth,clip]{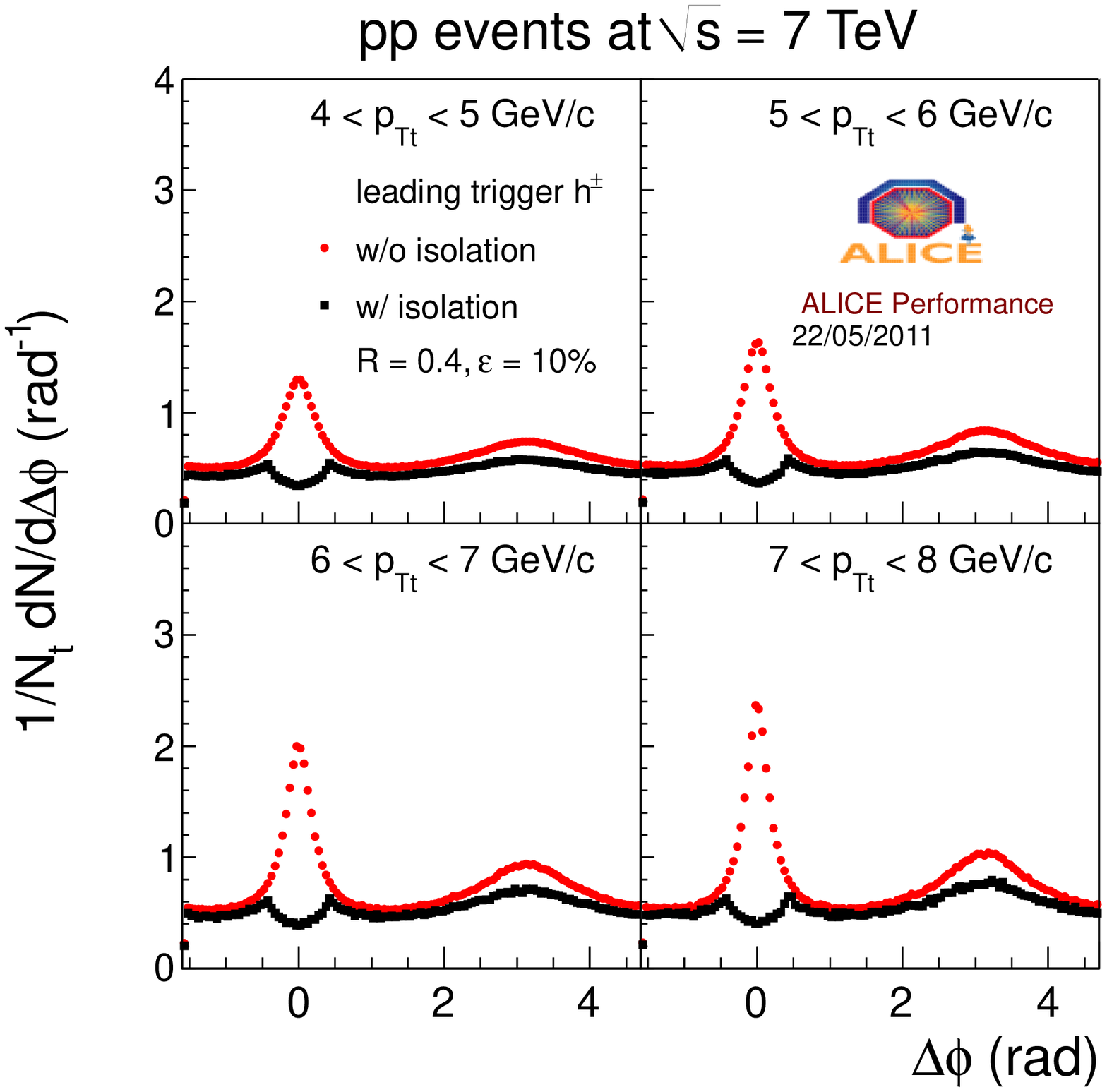}
\caption{\em Two-particle correlation distribution as a function of the relative azimuthal angle between the trigger and  associate charged particles  \dphi$=\phi_{\mathrm{t}}-\phi_{\mathrm{a}}$ for non-isolated and isolated charged triggers in pp collisions at \vs~= 7 TeV.}
\label{fig:IcTrigDphi}
\end{minipage}
\centering
\begin{minipage}{19pc}
\includegraphics[width=0.85\textwidth,clip]{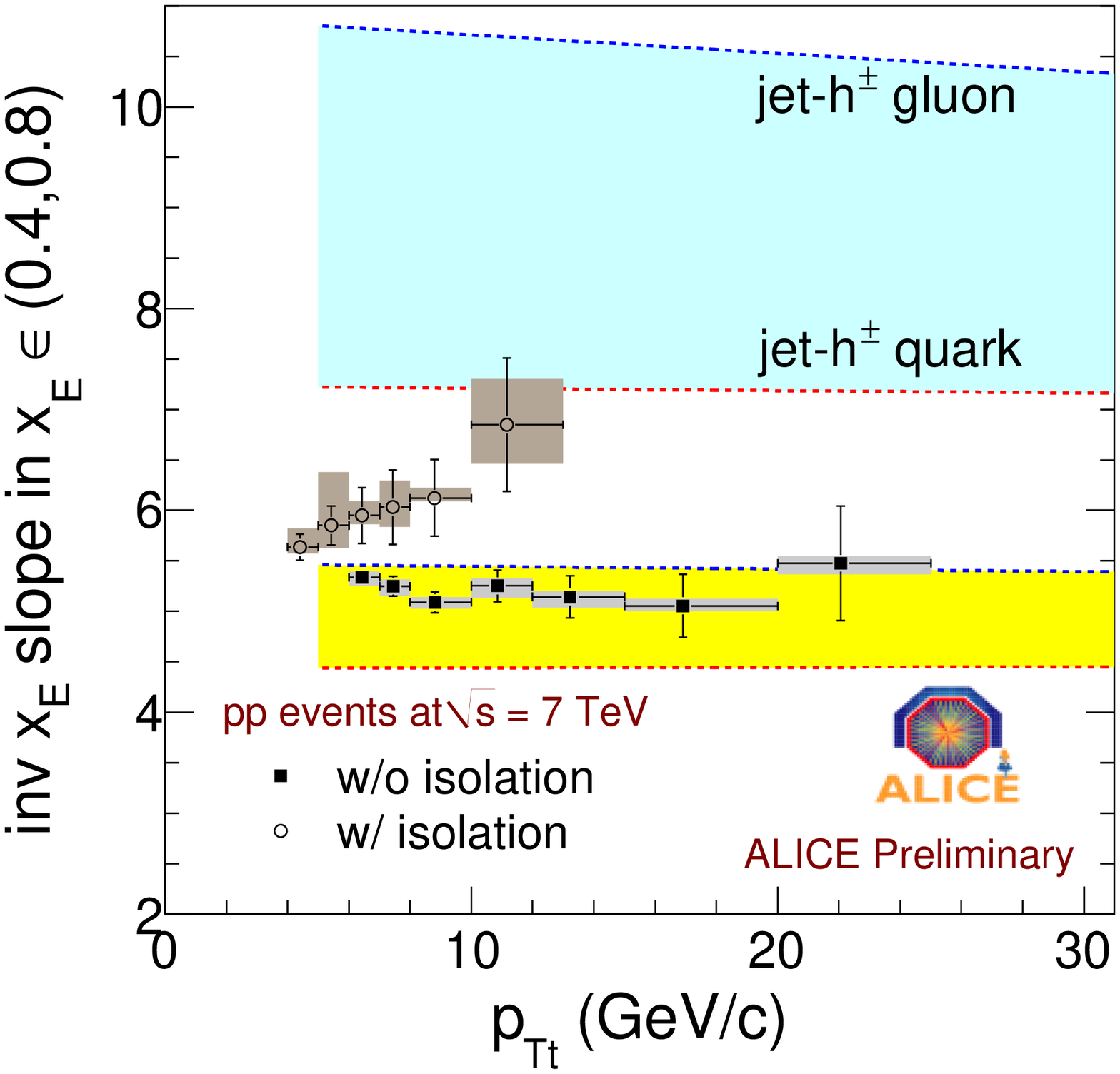}
\caption{\em Extracted inverse \xe slope as a function of the mean trigger \pt{t}~ for non-isolated and isolated charge triggers in pp collisions at \vs~= 7 TeV.}
\label{fig:IcSlope}
\end{minipage}
\end{figure}

Similar measurements performed at PHENIX~\cite{PHENIX2} conclude that the \xe distribution is not sensitive to the away side jet fragmentation function because the trigger particle carries only a small fraction of the full jet energy (\zt~= $\frac{p_{T_{t}}}{p_{T_{jet}}} < 1$). This conclusion is confirmed by our measurement where the measured inverse slope is in the region where one expects (yellow band) the \xe~slope when the trigger particle samples the jet momentum at \zt~$\sim 0.5$.
The next step of the analysis consists in selecting isolated trigger with the goal to enrich the trigger sample with single hadron jet events or direct photon  (\zt~$\rightarrow 1$). To select isolated particles, one studies the charged hadronic activity around the trigger candidate, i.e., calculating the sum of the transverse momentum of all charged hadrons inside a cone with radius $R = 0.4$ rad around the trigger candidate and tags the trigger to be isolated if the sum is less than 10~\% of the trigger's transverse momentum. 
The azimuthal correlation obtained with an isolated trigger (Fig.~\ref{fig:IcTrigDphi}), compared to the one without isolation, shows an apparent single jet structure with, by construction, the absence of the near side peak whereas the away side peak remains present. 
The slight difference in the away side peak intensity before and after isolation can be explained by a \zt  bias which is different in the case with and without isolation. From a Monte-Carlo study based on the PYTHIA event generator we have demonstrated that  non-isolated triggers carry on average 50\% of the jet energy while isolated triggers carry on average 80\% of the jet energy. 
Therefore, for the same \pt{} trigger, the isolated one will sample a lower energy jet compared to the non-isolated trigger.  
The resulting inverse \xe~slope parameter $n$ obtained by the exponential fitting for the non-isolated and isolated triggers are presented in Fig.~\ref{fig:IcSlope} as a function of the trigger transverse momentum. One observes that the $n$-value is generally higher for the isolated triggers than for the non-isolated triggers. 
The data points are also compared with prediction of a simple fragmentation model based on parent-child relation~\cite{Bjorken}.
The slope approaches the region where one expects $n$ to represent the slope of the true fragmentation function for quarks and gluons (blue band defined by \zt~$=1$).

The results demonstrated that two particle correlation is a suitable tool to study jet properties, where two particle correlations show a di-jet structure and the isolated trigger-hadron correlations make a better description of the away side jet fragmentation. 
The future study will focus on the identified particle correlations which has potential to distinguish the quark jets (direct photon trigger) and gluon jets (\piz trigger).

\end{document}